\documentclass[a4paper, fleqn]{article}
\usepackage[latin2]{inputenc}
\usepackage{ae}
\usepackage{aecompl}
\usepackage{aeguill}
\usepackage[british]{babel}
\usepackage[pdftex]{graphicx}
\usepackage[square, numbers, sort&compress]{natbib}
\usepackage[hyphens]{url}
\usepackage{geometry}
\usepackage{fourier}
\usepackage[dvipsnames, x11names]{xcolor}
\usepackage[pdftex, colorlinks = true, linkcolor = Maroon, citecolor = Maroon, urlcolor = PaleTurquoise4, pdfstartview = FitH]{hyperref}%
\begin{document}

\bibliographystyle{unsrt}

\author{Katalin Kopasz, Péter Makra and Zoltán Gingl}
%{Department of Experimental Physics, University of Szeged, Dóm tér 9, Szeged, Hungary, H6720}
\title{Edaq530: a transparent, open-end and open-source measurement solution in natural science education}
\date{}

\maketitle

\begin{abstract}

We present Edaq530, a low-cost, compact and easy-to-use digital measurement solution consisting of a thumb-sized USB-to-sensor interface and a measurement software. The solution is fully open-source, our aim being to provide a viable alternative to professional solutions. Our main focus in designing Edaq530 has been versatility and transparency. In this paper, we shall introduce the capabilities of Edaq530, complement it by showing a few sample experiments, and discuss the feedback we have received in the course of a teacher training workshop in which the participants received personal copies of Edaq530 and later made reports on how they could utilise Edaq530 in their teaching.

\vspace{5 pt}


\end{abstract}

\section{Introduction}

An article on natural science education in 2010 hardly needs to elaborate on how the use of computers can open up new horizons in demonstrating natural phenomena and the laws that describe them. Though many associate computer-aided education with simulations only, computers can also prove very efficient in enhancing real measurements. Just think of how more engaging and informative it can be to see the temperature change online on the screen during a calorimetric experiment than to reconstruct the whole process from three thermometer readings.

Consequently, the list of computer-aided education projects is longer than such an introduction could endeavour to cover in detail. In our view, these projects fall into one of three categories: professional solutions, household recipes and independent developments. Each of these categories can be viable, though they do target slightly different audiences.

Professional solutions (like \cite{Vernier-0, PicoTech-0, Butlin-0, Pasco-0, 0VirtIns, Bio-0}, for example) incorporate genuine measurement instruments, are developed and maintained by expert groups, offer official support and thus are the natural choice of many teachers. In many cases, though, adopting these solutions, especially in numbers required to outfit student laboratories, may be beyond the means of the school.

On the other end of the spectrum are household recipes \cite{Fence-0, Klaper-0, Courtney-0}, which focus on the innovative use of equipment already available to most people (the sound card, for instance). Though the creativity and the low cost of these ideas is appealing to many, they demand a high level of expertise and dedication to implement and maintain, making their wide-spread use in everyday classes highly unlikely. Furthermore, the household devices used here are not measurement devices, and thus --- though they can be invaluable in demonstrations --- careful analysis is required before we could accept the values they yield as measurement results. One may, for example, trust the time or frequency measurement results obtained with a sound card, but amplitude measurements with such a device are highly questionable methodically.

Independent developments \cite{ComLab-0, ComLab-1, 0ISES, Murovec-0, VenDASys-0} are somewhere in between. Though they cannot compete with professional solutions in terms of background infrastructure and product support, they are often developed by dedicated experts who also use them during their main field of activity, thus they have a more direct link between use and design in their favour. Since these developments are not profit-orientated, they have the potential to be open-source, increasing their flexibility and adaptability.

The main challenges these solutions must all face are how to be versatile whilst remaining essentially simple, and, even more importantly, how to keep the measurement hardware and the associated software form being perceived as a `black box' --- that is, to prevent the connexion between device and reality from becoming obfuscated.

In addition, it is hard to predict what makes a solution successful. Highly sophisticated, multi-functional professional suites may gather dust unused if their conception of handiness does not meet that of the user. Slight differences in logic and implementation may decide whether users will find the solution convenient. This is what gives us confidence that even though there has been a large number of successful solutions established in this field, introducing our own concept is not without benefit.

Here, we present Edaq530, a versatile, open-end and open-source experimental solution whose main focus is to make the computer-aided measurement as simple, precise and transparent as possible. Edaq530 comprises a thumb-size, micro\-con\-trol\-ler-driven USB-to-sensor interface and an open-source measurement software. In developing the hardware, our priorities were versatility and simplicity: we tried to find the broadest intersection of sensor architectures so that most sensors could be connected directly to the interface without the need for further transducer units. In order to promote transparency, we built the measurement software on a set of principles. First, we wanted to emphasise two essential concepts of modern instrumentation, sensors and calibration, so we brought them to the level of the user interface. Sensors are accessed not only through editable controls but also through pop-up calibration windows which illustrate the process of sensor calibration graphically. Second, we wanted to keep the software away from doing everything for the student, from presenting final data that no longer require their input. We wanted to find a balance and have the software do everything that is cumbersome but not instructive, whilst leaving room for further ideas and creative data processing.

The primary focus in developing Edaq530 was to serve the needs of the demonstration experiments of secondary-school physics teachers, but given the wide range of sensors that can be attached to it, Edaq530 can also be used for measurements in biology or chemistry. Furthermore, it is not limited to teachers' demonstrations, but can also be used by students with ease. The solution can also assist university measurements and being open-source offers the possibility of tailoring it to specific needs.

We first introduced our Edaq530 solution within the framework of a teacher training workshop, where a USB-to-sensor interface unit for each participant was incorporated into the enrolment fee. For this reason, low cost had to be a major aspect of the design. The training included practical laboratories in which the participants could learn the use of the hardware and the software. We also held optional classes for secondary-school students where they could carry out physics experiments using Edaq530. The feedback of both the teachers and the students is encouraging.

\section{Hardware}

\begin{figure}[h]
	\centering
		\includegraphics[width = 0.5\textwidth]{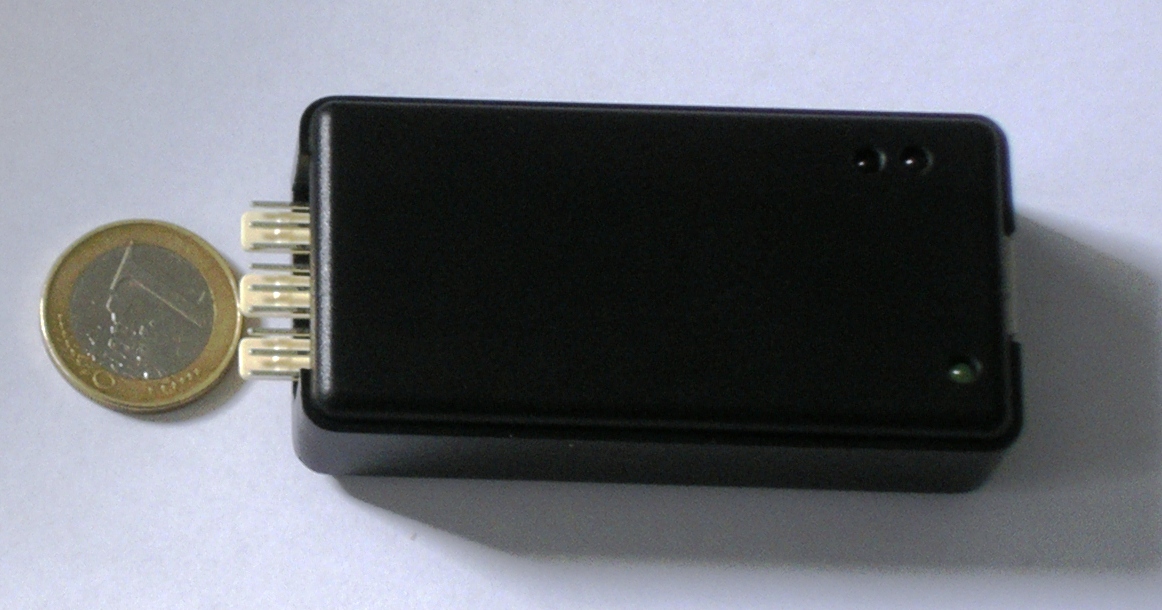}
	\caption{The Edaq530 USB-to-sensor interface.}
	\label{fig:Edaq530-0}
\end{figure}

The hardware unit of Edaq530 is a USB-to-sensor interface with dimensions of $4\,\mathrm{cm} \times 7.8\,\mathrm{cm} \times 2\,\mathrm{cm}$, so it can easily be carried in a pocket and there will still be room for a couple of sensors (see figure \ref{fig:Edaq530-0}). The device uses the power supply provided by the USB port, thus all it needs to be ready for measurement is a single USB cable connected to a computer. We have published the hardware documentation on our homepage (\url{http://www.noise.physx.u-szeged.hu/EduDev/EDAQ530/}) in order to make it freely reproducible and thus as `open-source' as the software.

Three sensors can be connected to Edaq530 simultaneously. The three sensor ports have identical three-pin connectors: the extreme pins correspond to the ground and the reference voltage, whilst the middle one carries the signal. The signal voltage must be between 0\,V and the reference voltage (3.3\,V for current models). We have tried to keep the sensor connexion as simple and versatile as possible, and found that this arrangement fits all the sensors we have adapted. The hardware includes internal pre-amplification for weaker signals (such as the signals of pressure sensors and thermocouples), internal probe resistances which can be switched on to provide resistance-to-voltage conversion for resistance-output sensors such as thermistors, and a differential measurement mode for Wheatstone-bridge sensors such as pressure sensors or load cells. The hardware units also contain a built-in infrared emitter-receiver pair that can function as a plethysmograph---drawing on the principle that the infrared transmission of a finger, for example, depends on the local haemoglobin concentration and is thus connected to instantaneous blood pressure--- or can be applied as a motion sensor.

\begin{figure}[h]
	\centering
		\includegraphics[width = 0.6\textwidth]{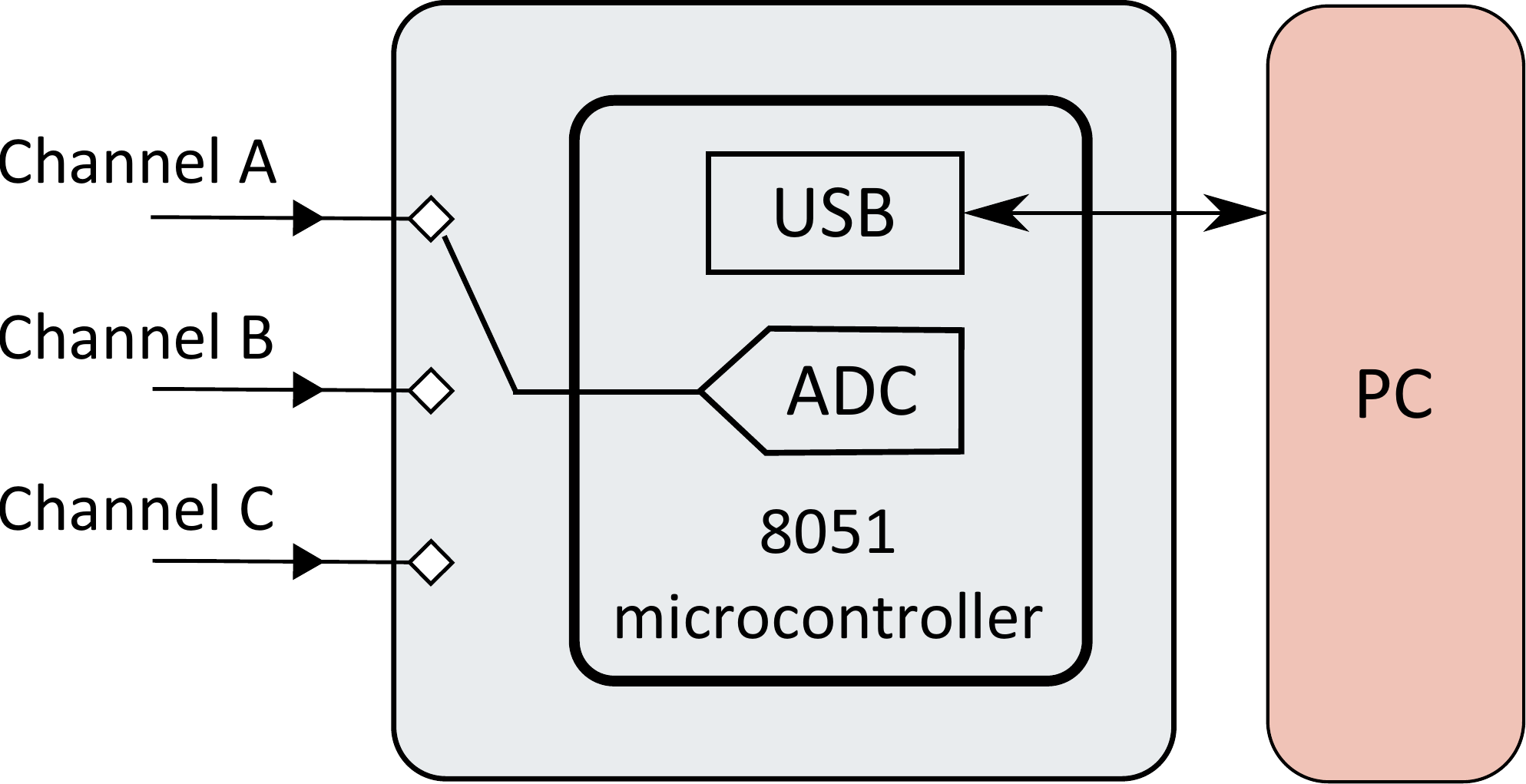}
	\caption{The block diagram of the Edaq530 hardware.}
	\label{fig:BlockDiagram-0}
\end{figure}

The hardware contains a single 12-bit analogue-to-digital converter unit and a three-channel multiplexer (see figure \ref{fig:BlockDiagram-0}). The device uses an internal sampling rate which is three times the sampling rate specified by the user, and performs the conversions on the three channels consecutively within the same external sampling period. The channels are consequently shifted relative to each other by one-third of the sampling interval. There is also the option of hardware averaging, for which the internal sampling rate is increased further and the device sends the average of multiple measurements to the computer at the rate specified by the user. The maximum of the sampling rate (the external sampling rate specified by the user, not the internal rate) is 1000\,Hz.

The core of the USB-to-sensor interface is a C8051F530A Silicon Laboratories microcontroller \cite{SiLabs-0} running our command interpreter program. The PC software sends simple text commands through the USB port, and the interface interprets and executes these commands (setting the sampling frequency, for example). This way a simple yet efficient communication can be established between the USB-to-sensor interface and the host computer.

\section{Software}

We developed the software in the .NET framework version 3.5 in C\# language. Although this fact ties it to the Windows platform somewhat, it has the potential to be ported to Linux or Mac operating systems through the Mono project. The program and its source code are available at \url{http://www.noise.physx.u-szeged.hu/EduDev/EDAQ530/}.

The software monitors the three channels of the Edaq530 hardware simultaneously. Each channel represents the data flow associated with a given connector on the hardware. The data are collected in the background, and can be displayed either on an online chart or in a numerical indicator. We decided against hiding the more `technical' measurement parameters such as sampling rate or averaging settings; the user has access to these and can tailor them to specific measurements. To turn raw voltages measured by the hardware into values of physical quantities, the user can associate a sensor with each channel independently. Sensors can be configured either by setting their parameters directly or through calibration. The program can perform level-crossing detection on every channel with channel-dependent parameters. Measurement settings, including sampling rate, time frame and the sensor information for all channels, can be saved into a stand-alone .xml file, facilitating the exchange of experiments between users.

\begin{figure}[h!]
	\centering
		\includegraphics[width = 0.49\textwidth]{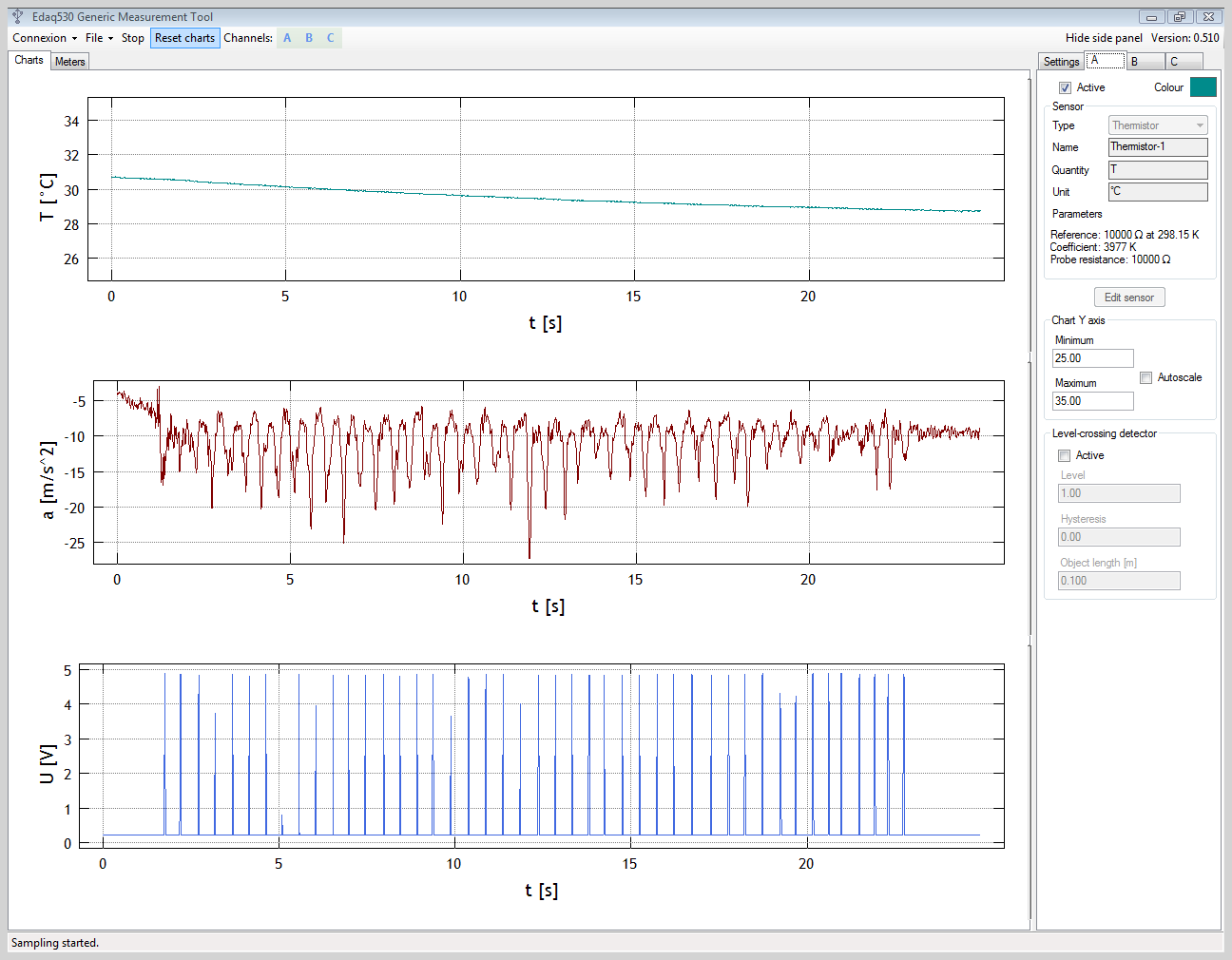}
		\includegraphics[width = 0.49\textwidth]{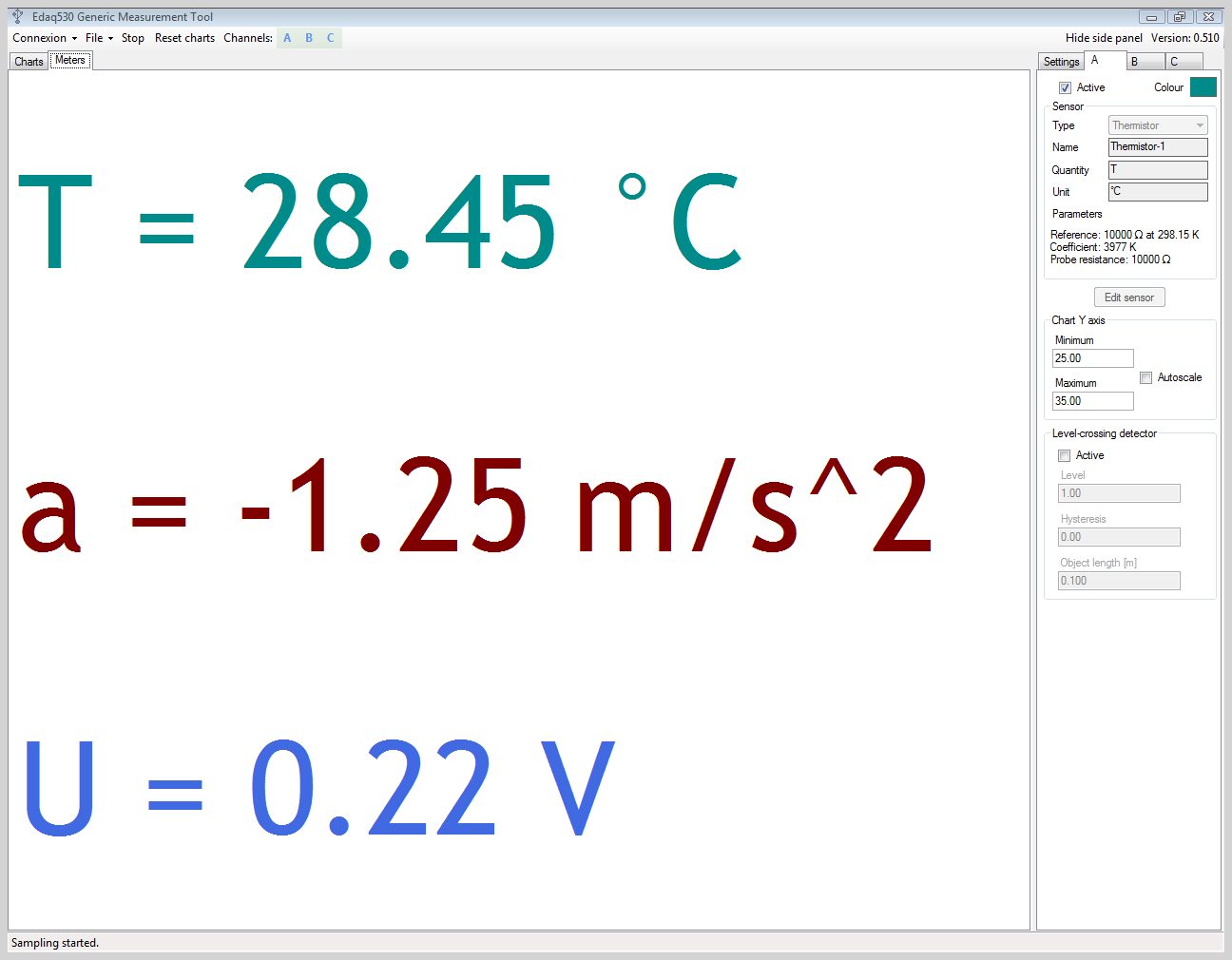}
		\caption{The software in chart mode (left panel) and in numeric display mode (right panel).}
	\label{fig:Chart-0}
\end{figure}

\subsection{Sensors}

Digital measurement devices apply analogue-to-digital converters to turn real signals into computer data. These analogue-to-digital converters can only take voltage as their input, yet rarely do we use a digital instrument as a mere voltmeter. In order to transform a given physical quantity into a voltage value that the analogue-to-digital converter can then process, we need a sensor.

The concept of sensors is essential in modern digital measurement technology. Sensors represent the key to the versatility of digital devices. For most physical quantities, one can find a sensor that translates the value of the quantity into a voltage. The relationship between quantity and voltage need not be linear; any kind of one-to-one relationship will do as long as we can find a feasible formula for it.

Our software representation of a sensor does the reverse of the sensor hardware: it takes the voltage supplied by an analogue-to-digital converter and translates it into the appropriate quantity on the basis of the sensor rule. We have implemented two types of sensor rules: linear scaling and thermistor scaling. These two types of scaling have sufficed to describe all sensors we have considered: thermistors, thermocouples, accelerometers, pressure sensors, etc. The output of the software representation of a sensor is the value that appears on the chart or the numeric indicator. If there is no sensor associated with a channel, the data in the channel is interpreted as voltage.

Linear sensors can be fully characterised by two parameters: the slope and the intercept. For the negative temperature coefficient thermistors we used, there are formulae with different order, but we have chosen a simple exponential approximation also given by two parameters: the resistance $R_0$ at the reference temperature $T_0$ and an exponential coefficient $B$. According to the formula we used, the resistance of a thermistor at temperature $T$ is

\begin{equation}
	R(T) = R_0 \mathrm{e}^{B\left(1/T - 1/T_0\right)}.\label{eq:thermistor}
\end{equation}

Of course, the hardware cannot measure resistance directly, so the the thermistor must first be placed into a voltage divider with a known probe resistance. Consequently, the scaling process of a thermistor is twofold: first, determining the thermistor resistance from the voltage across the thermistor, and then using the exponential thermistor rule in (\ref{eq:thermistor}) to find the temperature.

To emphasise the role of sensors in measurements, we have designed a sensor control and made it the central element of channel settings. Although the primary purpose of the sensor control is to set the scaling in a given channel, this is also where the user can enter the name and the unit displayed in the channel, and the changes submitted here are also reflected both in the title of the y-axis and on the numeric display. We also wanted to instil into students a sense of how a physical quantity is inseparable from its unit; for this purpose, we took great pains to display the unit of the slope and of the intercept together with its value.

\begin{figure}[h]
	\centering
		\includegraphics[width = 0.7\textwidth]{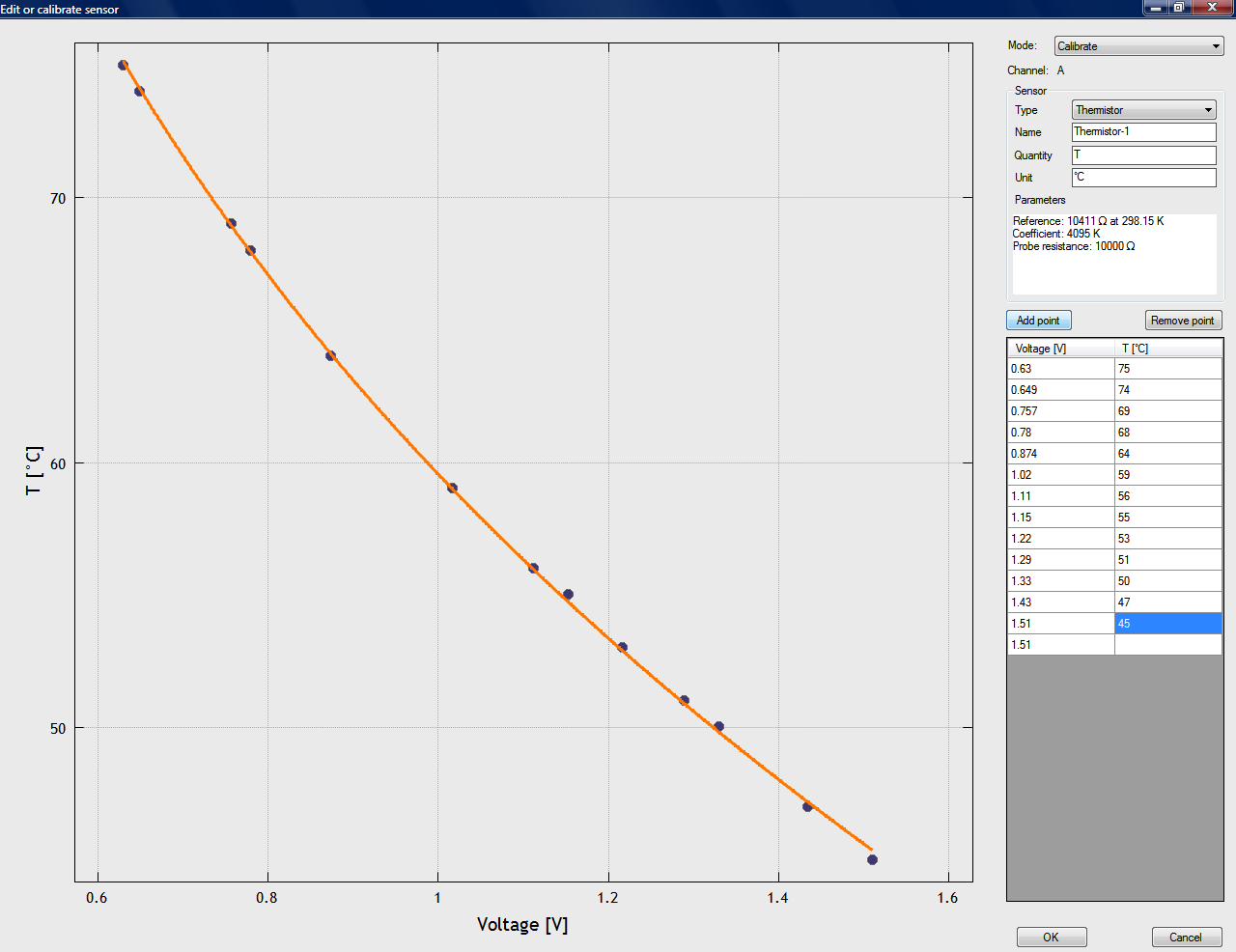}
		\includegraphics[width = 0.225\textwidth]{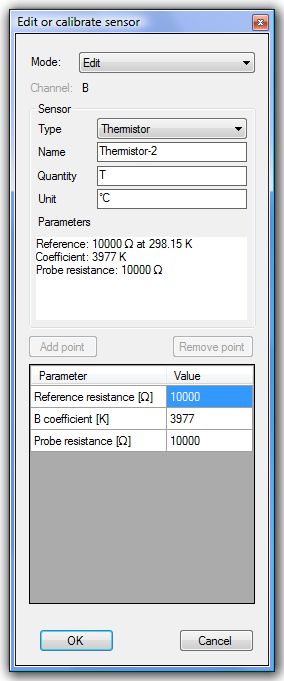}
	\caption{The sensor editor in calibration mode (left panel) and in edit mode (right panel).}
	\label{fig:Calibration}
\end{figure}

The parameters of sensors can be set in a sensor editor pop-up window (see figure \ref{fig:Calibration}). The sensor editor has two modes: in the editor mode, the user can enter the parameter values directly (eg, from the data sheet of the sensor), whilst in calibration mode, the device monitors the voltage in a channel, to which the user can assign the quantity value provided by an independent calibration device (eg, the temperature shown by a mercury thermometer). In calibration mode, the voltage-quantity pairs are stored and displayed in a table, as well as plotted in a graph along with the theoretical fit. The program performs a linear least squares fit to find the scaling parameters; for linear sensors, this is straightforward, but after linearising (\ref{eq:thermistor}), we can use the very same algorithm for thermistors as well. Whenever the user enters a new voltage-quantity pair, the program repeats the fit with the new data included and updates the graph, so outliers can be detected and removed from the table, if necessary. Apart from being handy in measurements, this feature can also familiarise students with another important measurement concept, calibration.

Once sensor parameters have been specified, sensor data can be saved into an .xml file to be reused later or shared between users.

\subsection{Level-crossing detection}

Most real-world events can be tied to a physical quantity crossing a threshold in either positive or negative direction. A well-known example could the photogate: when an air-track glider passes between the arms of a photogate, the voltage of the photogate will rise the instant the glider starts to block the path of the infrared beam, so if we compare photogate voltage to a level somewhere between the minimum corresponding to the unblocked beam and the maximum corresponding to the blocked beam, we can tell exactly when the glider passed through the gate. The use of level crossing detection is not limited to photogate signals, though: consider, for instance, the acceleration signal of a pendulum, where the distance between two neighbouring crossings of any non-zero level below the amplitude that have the same direction (that is, downwards or upwards) yields the period of the motion directly.

A natural field of application of time-instant detection is kinematics: knowing the positions of photogates and determining the time instants at which the gliders occupied these positions allows kinematic quantities to be estimated. Level crossings can also yield information on the swing time of a pendulum whose bob passes between the arms of a photogate (or to which we attach an accelerometer), or when identified within a thermistor signal, they can be used to determine the cooling rate, and with level crossing detection on the signal of an infrared plethysmograph, we can obtain even an instant heart rate sequence.

The software represents events with level crossing cycles. A cycle starts with an upwards crossing, stores the subsequent downwards crossing and ends with the next upwards crossing (which is also the beginning of the next cycle). Events which are tied to the negative slope of a signal can be made to fit this scheme by using a scaling which inverts the signal. The time span between two neighbouring upwards crossings carries information on the period of periodic motion, whilst the time elapsed between an upwards crossing and a subsequent downwards crossing can be used to determine the average speed of an object passing through a photogate: if we attach an object of well-defined length (such as a flag) to the glider, the upwards crossing is the time instance when the front edge of the flag intersects the infrared beam and the downwards crossing is the corresponding time instance for the back edge of the flag; dividing the length of the flag by the upwards-to-downwards time span will yield the average speed of the glider for the time interval of passing through the photogate. If the flag length is sufficiently short, this can yield a good approximation of the instantaneous speed.

\begin{figure}[h]
	\centering
		\includegraphics[width = 0.50\textwidth]{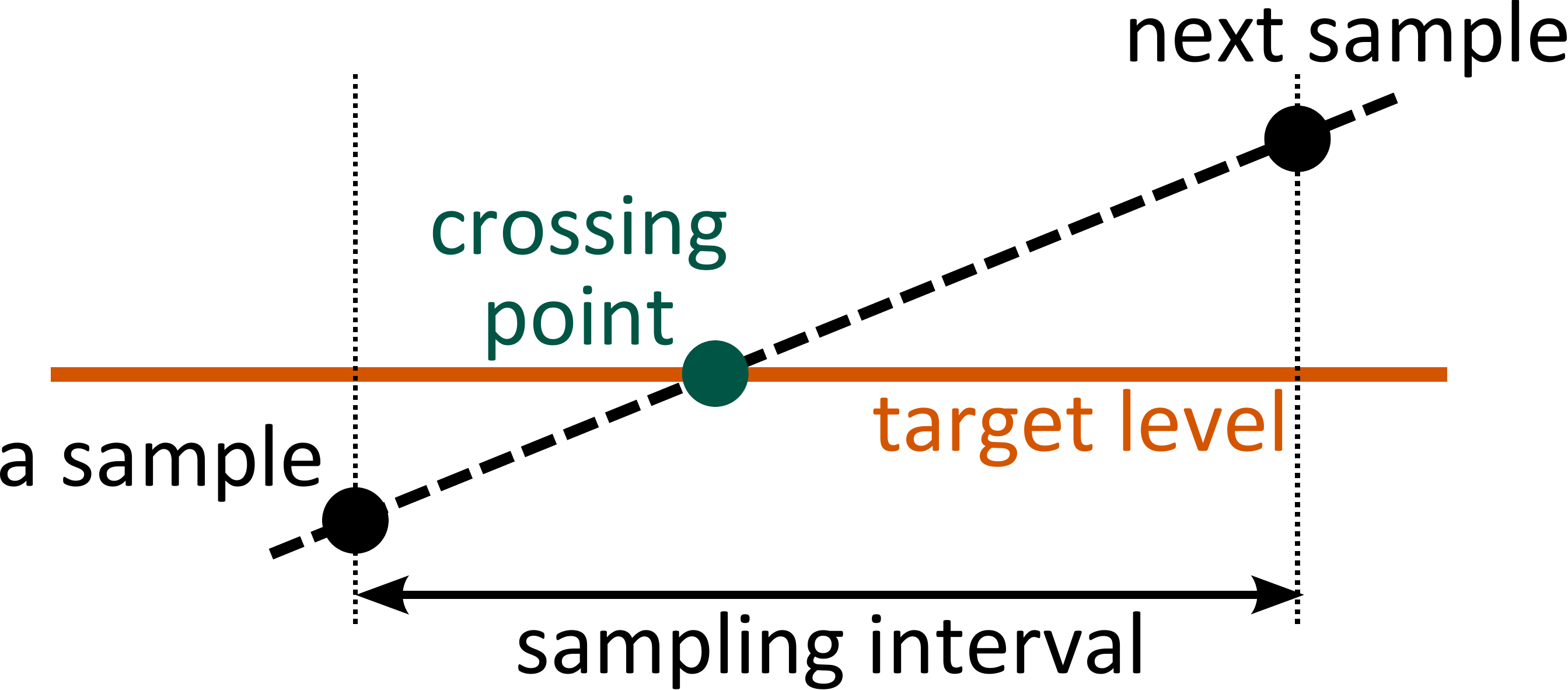}
	\caption{Interpolation improves time resolution of level crossing detection.}
	\label{fig:LevelCrossingScheme-0}
\end{figure}

In a simple level-crossing scheme, one must find a pair of consecutive points in the signal which are on opposite sides of the level to obtain a level crossing. The crossing itself is the time instance at which the linear interpolating function connecting these two points assumes the value of the level. Due to interpolation, the time resolution of the level-crossing detection is always better than the time resolution of the signal itself (see figure \ref{fig:LevelCrossingScheme-0}).

\begin{figure}[h]
	\centering
		\includegraphics[width = 0.7\textwidth]{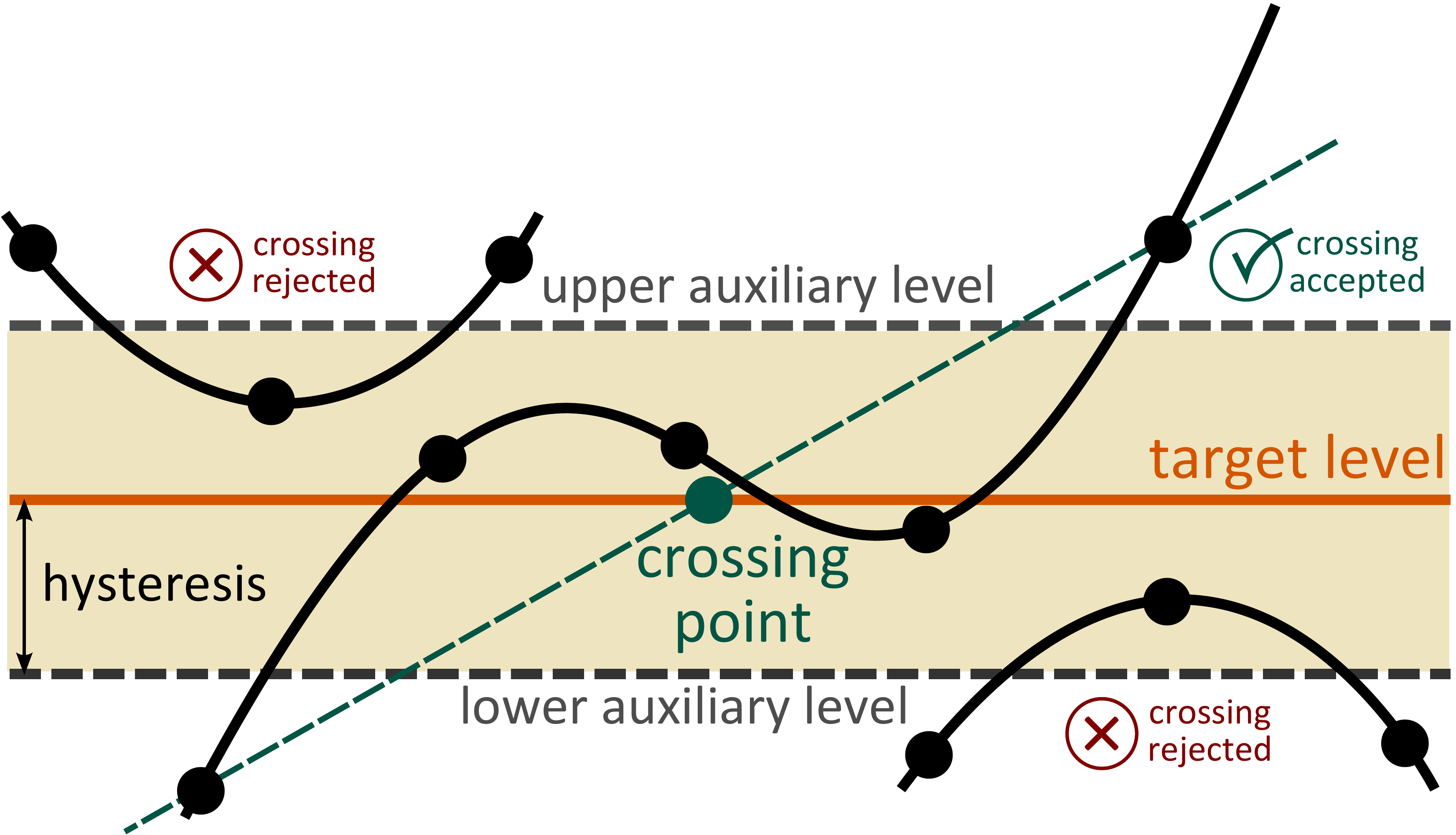}
	\caption{Level crossing detection using hysteresis.}
	\label{fig:LevelCrossingScheme-1}
\end{figure}

For noisy signals, this simple detection method is prone to report spurious level crossings when the noise causes the signal meander back to the other side of the level immediately after a crossing. To tackle this problem, we have adapted the idea applied in thermostatisation, for example, and introduced hysteresis into the level-crossing scheme. A thermostat uses two levels instead of one: one below the specified temperature and another above it. When the temperature drops below the lower level, heating starts, whilst when it rises above the upper level, heating stops. This way the system can be prevented from switching on and off constantly. Similarly, in our level crossing detection algorithm, we have two auxiliary levels symmetrically above and below the target level. We define the hysteresis parameter as the difference between the target level and one of the auxiliary levels; a hysteresis of 0 is identical to the simple detection scheme. Whenever we find two consecutive points in the signal which are on the opposite sides of an auxiliary level, we mark the first point as a pre-crossing candidate. If later the signal leaves the interval between the two auxiliary levels before crossing the other auxiliary level, the pre-crossing candidate is discarded and the search for pre-crossing candidates starts anew, but if the signal proceeds to cross the opposite auxiliary level without leaving this interval, we accept the pre-crossing candidate and mark the point immediately after the opposite crossing as a post-crossing point. We then determine the crossing as the time instance at which the linear interpolating function connecting the pre-crossing point with the post-crossing point takes the value given by the target level. This level crossing detection method is illustrated in figure \ref{fig:LevelCrossingScheme-1}.

This level crossing detection can be applied independently in any of the three channels or in all of them simultaneously. The data of the level-crossing cycles (the time instant of the leading upwards crossing, the period up to the next cycle and the speed calculated from object length and upwards-to-downwards time span) are displayed in a table and can be copied to the clipboard for processing later.

\subsection{Measurement templates}

Teachers often have a very limited amount of time to arrange demonstration experiments in class. It is therefore crucial for the measurement software to be able to load the measurement settings appropriate for an experiment (which have been prepared beforehand) or switch between the settings of experiments with a single click. Our program stores all settings relevant to an experiment in a single .xml file. The settings in a measurement template include sampling rate, averaging information, time window length, scale titles, the appearance of graphs (colours and displayed ranges), the active or inactive states of channels, the sensors and the level-crossing settings associated with channels. Once the appropriate settings have been found, the user can save the measurement template to be reused later, and they can switch between different settings without restarting the software. Measurement templates can also be shared between teachers.

\section{Sample experiments}

\subsection{Picket fence}

In a `picket fence' experiment \cite{Fence-0}, we drop a thin, elongated object with regularly alternating transmissive and less transmissive regions along its length (such as a plastic ruler taped over with adhesive tape at regular intervals) between the arms of a photogate. The photogate detects the time instants at which the beginning or the end of a less transmissive region crossed the infrared beam, and, knowing the distances between less transmissive regions, we can reconstruct the path of the free fall of the picket fence and thus demonstrate different aspects of linear motion with constant acceleration.

With level-crossing detection switched on, the Edaq530 software presents a table whose rows represent consecutive stripes on the picket fence crossing the beam of the photogate, the first column contains the time instant at which the beginning of the given stripe reached the photogate, the second displays the time span up to the next crossing and the third represents the average speed of the picket fence for the duration of the given stripe passing through the gate, calculated from stripe width and the time elapsed between the instant the beginning of the stripe crosses the beam and the subsequent crossing that signals the end of the stripe arriving there.

\begin{figure}[h]
	\centering
		\includegraphics[width=0.75\textwidth]{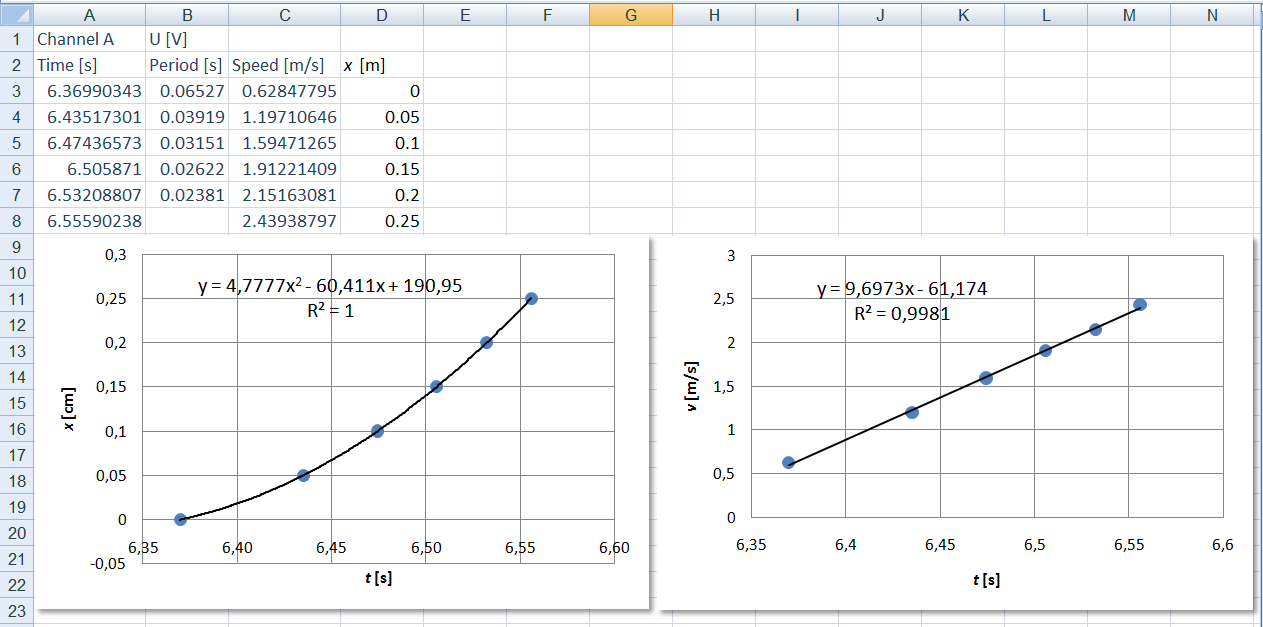}
	\caption{Picket fence experiment --- processing the data of Edaq530 in a spreadsheet program.}
	\label{fig:PicketFence-0}
\end{figure}

Having obtained the level-crossing data, we can copy them to the clipboard and paste them into a spreadsheet program. Plotting speeds as a function of time, we can demonstrate how the speed increases linearly with time, and the curve fitting function available to most spreadsheet programs can instantly yield an estimate of $g$ (see figure \ref{fig:PicketFence-0}). With an advanced group of students, we can analyse the error introduced by using the average speed during crossing intervals as an approximation of the instantaneous speed. The effects of the finite time resolution of Edaq530 can also be brought to attention.

If we insert into the spreadsheet a column containing the positions of stripe edges, we can also present the quadratic displacement--time graph of linear motion with constant acceleration. Applying a quadratic fit yields an estimate of $g/2$.

\subsection{Heat conduction}

\begin{figure}[h!]
	\centering
		\includegraphics[width = 0.55\textwidth]{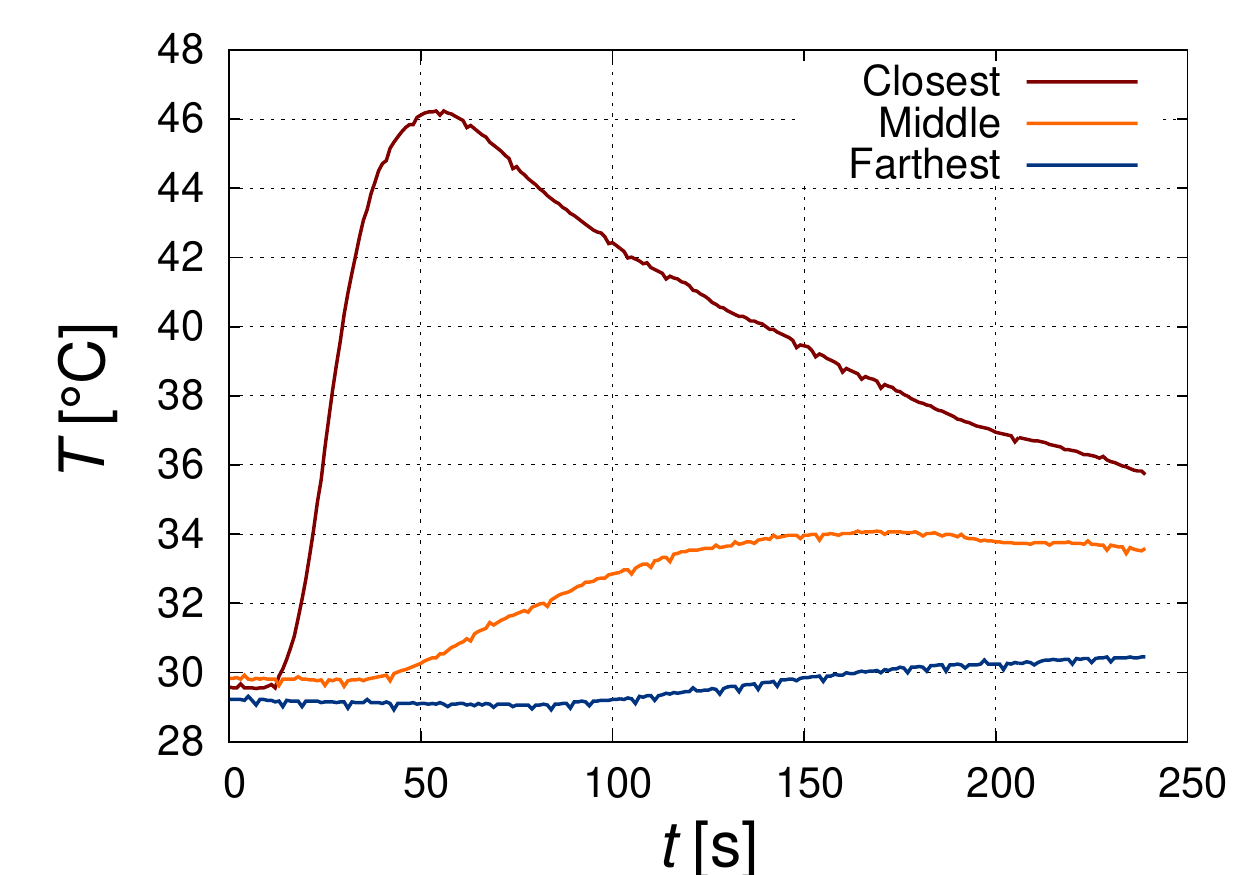}
	\caption{Temperatures of three thermistors along a metal rod. `Closest', `middle' and `farthest' are defined relative to the heating point.}
	\label{fig:HeatConduction-0}
\end{figure}

The phenomenon of heat conduction is hard to demonstrate using `classical' methods, but digital arrangements can make its study rather illustrative. We take a metal rod with a heat tank at one end, and place three thermistors at equal distances along its length. After applying a match or a lighter to one end, we can follow how heat propagates along the rod. Comparing the three thermistor signals, we can demonstrate that it takes time for heat to travel and the temperature peaks diminish in the process (see figure \ref{fig:HeatConduction-0}).

\subsection{Oscillations on a spring}

\begin{figure}[h!]
	\centering
		\includegraphics[width = 0.7\textwidth]{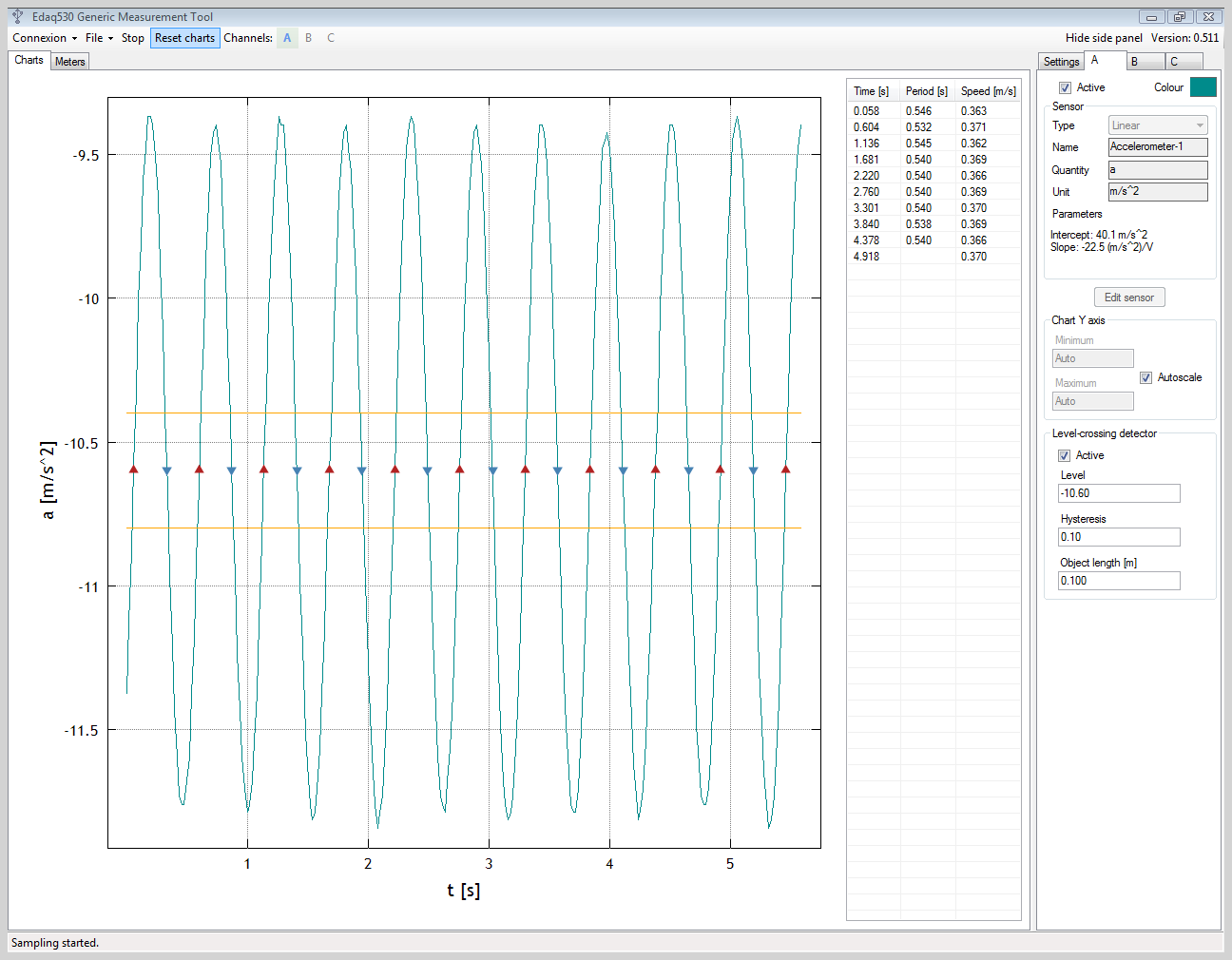}
	\caption{Oscillations on a spring.}
	\label{fig:Spring-0}
\end{figure}

Attaching an accelerometer to an object on a spring, we can illustrate harmonic oscillations. The software displays the acceleration versus time plot online, and, switching level-crossing detection on, we can instantly read the period of the motion (see figure \ref{fig:Spring-0}). If we set the object in motion so that it also has an initial displacement from the vertical position, the coupling between vertical oscillations and sideways pendulum motion can also be demonstrated. For the purposes of showing how the amplitude changes during oscillations and determining the period with level-crossing detection, the accelerometer can be replaced by other sensors such as a Hall sensor (with a small magnet attached to the oscillating object) or the built-in photosensor. The latter offers the most sensitive detection of motion of all the sensors we have used with Edaq530.

\section{Reception and feedback}

We tailored the hardware to a teacher training workshop that took place on 20 and 21 March 2010. We balanced performance and cost so that the price of a device could be incorporated into the enrolment fee. This way we hoped to embed our solution in the repertoire of the teacher community by offering each participant a personal copy of the hardware. The focus of the workshop was multimedia in education, and amongst lectures and seminars on utilising digital devices and techniques in teaching biology, chemistry and physics, the programme also included our talk introducing the Edaq530 solution. More importantly from our point of view, a significant part of the workshop was devoted to laboratory practice in which the participants could learn the use and the capabilities of Edaq530.

In order to get official credit for the teacher training workshop, participants had to render an account on what they had learnt during the workshop. They could choose between preparing a PowerPoint (or equivalent) presentation incorporating the techniques from the presentation-making seminar of the workshop and making a report on a demonstration experiment which uses Edaq530. 9 out of 11 participants chose the latter. The experiments spanned several areas of physics from the study of pendulum motion, circular motion or the conservation of linear momentum to the investigation of the effectiveness of thermal insulation. The participants addressed different aspects of teaching in which Edaq530 can assist them. One participant related how seeing the effects of actions (like touching a thermistor or moving an accelerometer) can help students establish or reinforce the concept of Cartesian graphs as representations of real-world events. Another participant saw the potential of the Edaq530 solution in presenting measurement proof for laws or principles like the conservation of linear momentum, complemented by the quantification of the difference between empirical data and theoretical expectations and an analysis of the causes of the difference. There was a presentation which also reflected on the limitations of the accelerometer applied (such as finite response time and limited measurement range), which could introduce students already orientated towards natural sciences to the intricacies of real measurements.

In general, our impression was that Edaq530 served the creativity of the teachers well: they could use the same simple software for a wide range of assorted experiments, and on the basis of the feedback provided by the reports, we feel that the comfort functions like built-in level-crossing detection (as compared to having to read a cursor like in PicoLog \cite{0PicoLog}) can indeed speed up the measurements so that their inclusion in a very limited classroom time is viable. Furthermore, we perceived that using the software had raised in the participants the level of awareness of the essential aspects of measurements and the pitfalls one should try to avoid, which seemed to have justified our decision not to hide `technical' measurement parameters such as sampling rate. One of the reports has shown us that teachers see potential also in using the device in educating students with less aptitude for natural sciences.

The participants also filled in a survey form about their evaluation of the programme. One of the questions of the survey asked the participants to assess to what extent they think they would be able to incorporate the workshop material into their teaching routine. Eight out of 11 answers fell into the category between 75\% and 100\%.

Parallel to teacher training, we also tested Edaq530 in an optional class for secondary-school students interested in physics. In the course of the class, students carried out physics experiments with Edaq530. Our experiences have shown that students can learn the use of the device with ease, and they are motivated to be creative in their experiments, coming up with ideas we had not considered before. One of the students' ideas was, for example, to apply the photogate to examine the transmission of the plastic ruler functioning as a picket fence.

\section{Conclusion and prospects}

In this paper, we introduced our Edaq530 measurement system and argued for its potential as a viable open-source addition to the existing solutions in natural science education. The main argument to support our claim is that we have had the opportunity to test---both with students and with teachers---whether our design principles of transparency and a balance of simplicity and versatility work in practice, and the feedback we have received is decidedly positive. We have reason to think that our Edaq530 solution can successfully be integrated into secondary education, helping teachers demonstrate concepts of natural sciences whilst motivating students to appropriate these concepts and refocussing their attention onto natural sciences. In addition to secondary education, it can also serve university-level experimentation as a flexible measurement tool.

We are participating in an educational project to furnish a model laboratory with Edaq530 units along with a wide range of sensors. The laboratory is hosted by a secondary school, and optional classes in the laboratory started in the autumn of 2010 with 20 measurement sites. We hope that apart from advancing the physics education of talented students, these classes will also provide us with valuable feedback as how to improve the Edaq530 solution. Due to the low cost of the device, lending it to students for home experimentation is also an option. We continue to develop new sensor architectures and software updates that reflect hardware changes or user requests. One of the development options we are currently exploring is Bluetooth connectivity.

\section*{Acknowledgements}

This work was supported by the Hungarian National Research Fund (OTKA K-69018).

\bibliography{Edaq530}

\appendix
\section*{Appendix: The hardware}

The detailed documentation of the hardware is available at \url{http://www.noise.physx.u-szeged.hu/EduDev/EDAQ530/}.

\begin{figure}[h]
	\centering
		\includegraphics[width=1.00\textwidth]{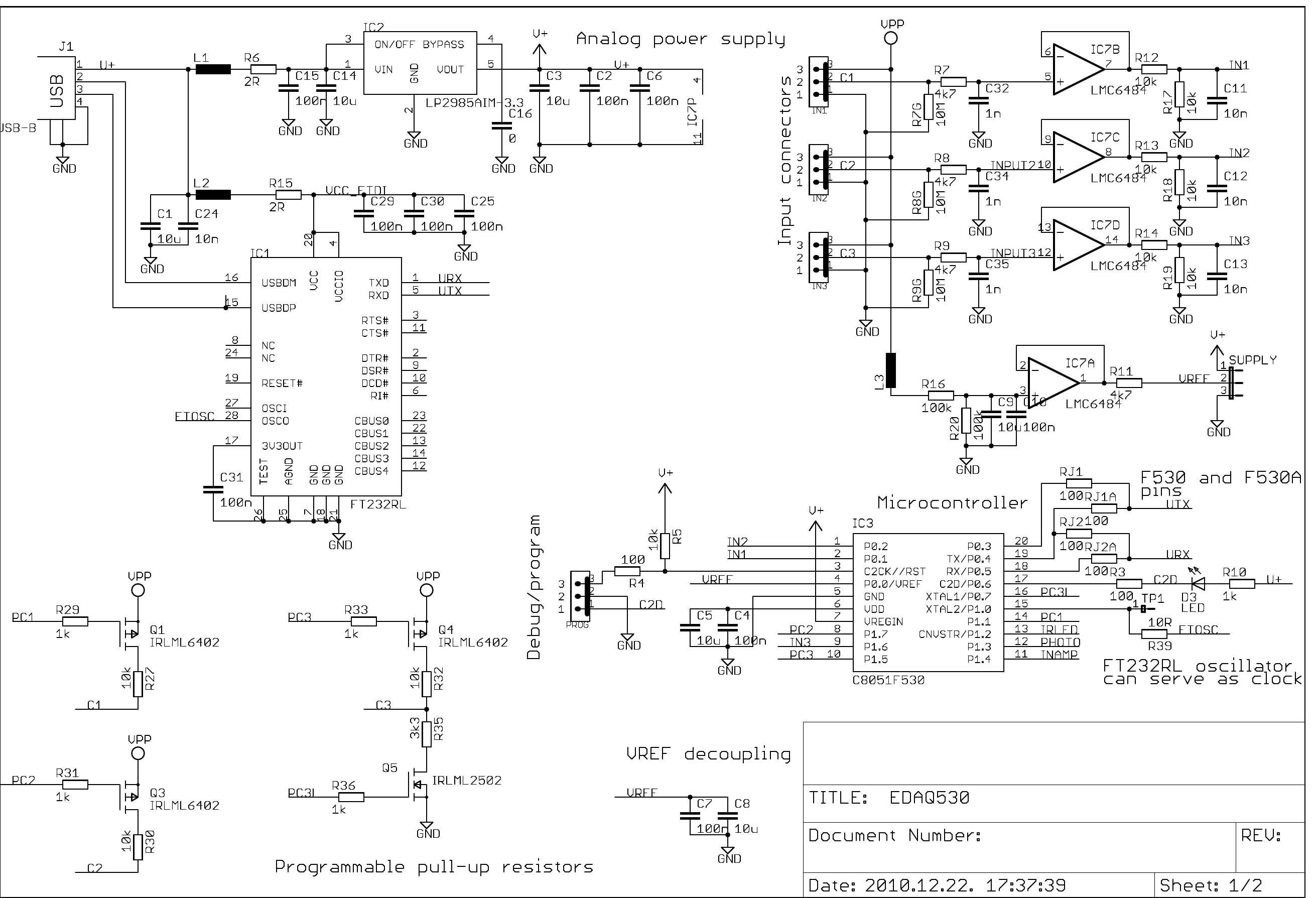}
	\caption{The schematic diagram of the Edaq530 circuit board, page 1.}
	\label{fig:Fig10}
\end{figure}

\begin{figure}[h]
	\centering
		\includegraphics[width=1.00\textwidth]{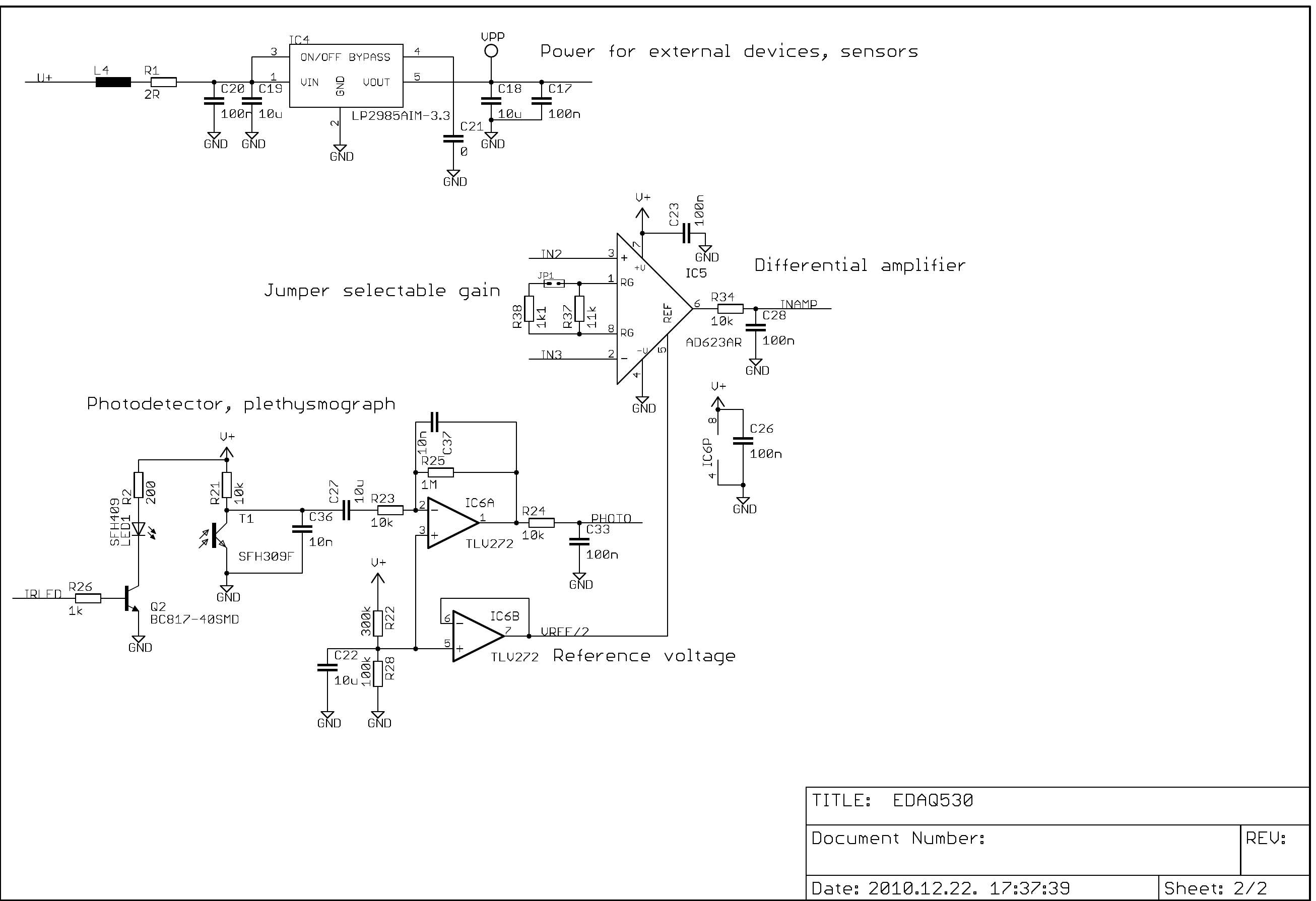}
	\caption{The schematic diagram of the Edaq530 circuit board, page 2.}
	\label{fig:Fig10}
\end{figure}

\begin{figure}[h]
	\centering
		\includegraphics[width = 0.8\textwidth]{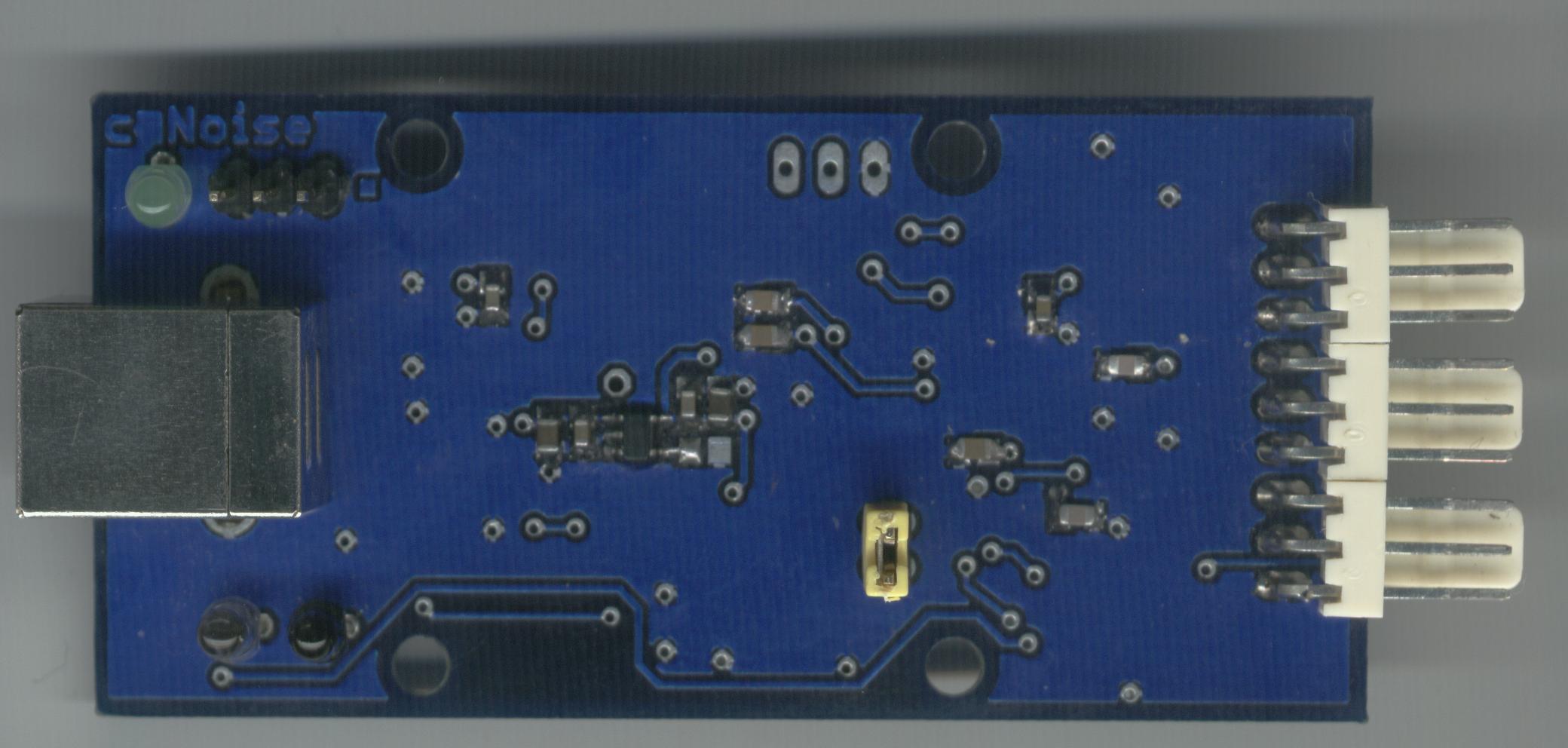}
	\caption{The top of the Edaq530 circuit board.}
	\label{fig:Fig10}
\end{figure}

\begin{figure}[h]
	\centering
		\includegraphics[width = 0.8\textwidth]{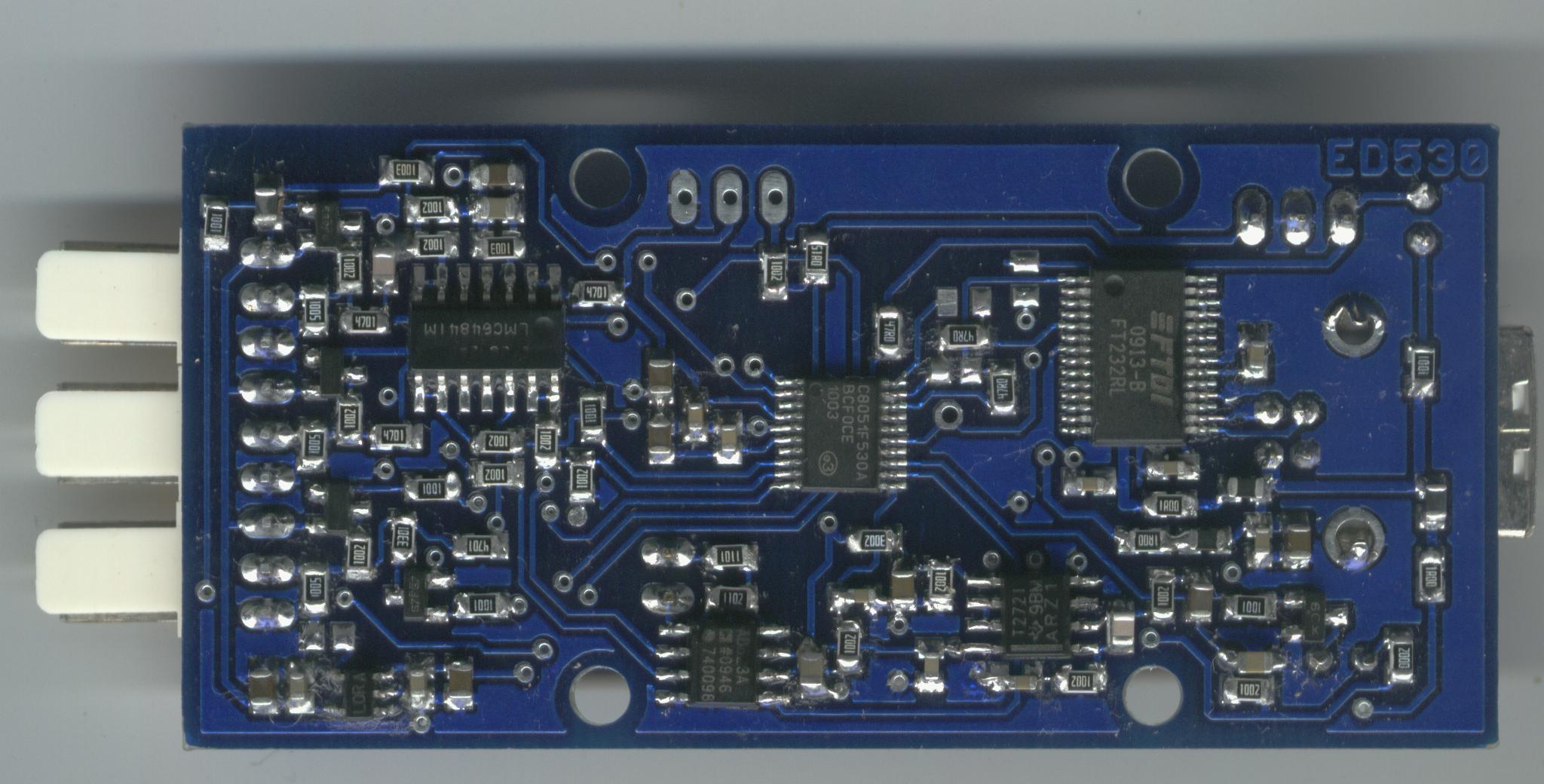}
	\caption{The bottom of the Edaq530 circuit board.}
	\label{fig:Fig10}
\end{figure}

\end{document}